\documentclass[aps,twocolumn,amssymb,showpacs]{revtex4}

\usepackage{amsmath}
\usepackage{amssymb}

\renewcommand{\(}{\left(}
\renewcommand{\)}{\right)}
\renewcommand{\[}{\left[}
\renewcommand{\]}{\right]}

\newcommand{\D}{\nabla}
\renewcommand{\d}{\partial}

\newcommand{\beq}{\begin{equation}}
\newcommand{\eeq}{\end{equation}}
\newcommand{\beqa}{\begin{eqnarray}}
\newcommand{\eeqa}{\end{eqnarray}}
\newcommand{\nol}{\nonumber}

\begin{document}

\title{More on the covariant retarded Green's function for the electromagnetic field in de Sitter spacetime}

\author{Atsushi HIGUCHI}
\email{ah28@york.ac.uk}
\author{LEE Yen Cheong}
\email{yl538@york.ac.uk}
\author{Jack R.~NICHOLAS}
\email{jacknicholas87@googlemail.com}
\affiliation{Department of Mathematics, University of York,
Heslington, York YO10 5DD, United Kingdom}

\date{July 22, 2009}

\begin{abstract}

In a recent paper~\cite{Higuchi:2008fu} it was shown in examples that the covariant retarded Green's
functions in particular gauges for electromagnetism and linearized gravity
can be used to reproduce field configurations correctly
in spite of the spacelike nature of past infinity in de Sitter spacetime.
In this paper we extend the work of Ref.~\cite{Higuchi:2008fu} concerning the electromagnetic field
and show that the covariant retarded Green's function with an arbitrary value of the
gauge parameter reproduces the electromagnetic field from two opposite charges
at antipodal points of de Sitter spacetime.

\end{abstract}


\maketitle

It has been claimed by some authors that the spacelike nature of past infinity of de Sitter spacetime
invalidates the covariant retarded Green's functions
in electromagnetism and linearized gravity in de~Sitter spacetime (see, e.g.~Ref.~\cite{Woodard}).
As was pointed out in Ref.~\cite{Higuchi:2008fu}, in a spacetime with spacelike past infinity, such as
de~Sitter spacetime, one needs to take into account the contribution from the initial data on past infinity,
which are necessarily nonzero if the charge density does not vanish there~\cite{Philosophy,Bicak}, when calculating
the field configuration using the retarded Green's function.
This issue does not arise in Minkowski spacetime, and that is why some authors
consider only the contribution from the charge and erroneously conclude that the retarded Green's function does not
work in de~Sitter spacetime.

In Ref.~\cite{Higuchi:2008fu} one example for showing how the retarded Green's function should be used
in de~Sitter spacetime was the
electromagnetic field produced by two charges with opposite signs placed at antipodal points.   This example
was studied in the Feynman gauge.  The purpose of this paper is to extend this example by
using the covariant gauge with an arbitrary value of the gauge parameter.

We consider the free electromagnetic field described by the Lagrangian
\beq
\mathcal{L} = \sqrt{-g}\left[- \frac{1}{4}F_{ab}F^{ab} - \frac{1}{2\zeta}(\nabla_a A^a)^2\right]\,,
\eeq
where $g$ is the determinant of the background de~Sitter metric and where
$F_{ab} = \nabla_a A_b - \nabla_b A_a$.
Let $J^a(x)$ be a current coupled to the electromagnetic potential $A_a(x)$ and let $\Sigma$ be a Cauchy
surface.  Then, as was explained in Ref.~\cite{Higuchi:2008fu},
the field $A_a$ in the future domain of dependence of $\Sigma$, denoted $D^+(\Sigma)$, is given in
terms of the retarded Green's function $G_{ab'}(x,x')$ by
\beq
A_a(x) = A_a^{(S)}(x) + A_a^{(I)}(x)\,, \label{AHreproduce}
\eeq
where
\beqa
A_a^{(S)}(x) & = & \int_{D^+(\Sigma)}d^4 x'\sqrt{-g(x')}G_{ab'}(x,x')J^{b'}(x')\,,\label{AHsource}\\
A_a^{(I)}(x) & = & \int_{\Sigma}d\Sigma_{a'}\left[\pi^{a'b'}(x')G_{ab'}(x,x')\right.\nonumber \\
&& \left. - A_{b'}(x'){(L_\pi G)_{a}}^{a'b'}(x,x')\right]\,, \label{AHinitial}
\eeqa
with
\beqa
\pi^{a'b'} & = & \frac{1}{\sqrt{-g}}\frac{\partial\mathcal{L}}{\partial(\nabla_{a'} A_{b'})}\nonumber \\
& = & -F^{a'b'} - \zeta^{-1}g^{a'b'}\nabla_{c'} A^{c'}\,, \label{AHpi}\\
{(L_{\pi}\tilde{G})_a}^{a'b'} & = & - 2\D^{\left[a'\right.}\tilde{G}_a{}^{\left.\ b'\right]}
- \zeta^{-1}g^{a'b'}\D^{c'}\tilde{G}_{ac'}
    \,. \label{LG}
\eeqa
As in Ref.~\cite{Higuchi:2008fu} we call the fields $A_a^{(S)}(x)$ and $A_a^{(I)}(x)$ the \emph{source field}
and \emph{initial field}, respectively.  In de~Sitter spacetime the initial field with past infinity adopted as $\Sigma$
is crucial in reproducing the field in terms of the retarded Green's function. This fact was overlooked in the
erroneous claim that the retarded Green's function in the covariant gauge is invalid in de~Sitter spacetime.

The metric of de~Sitter spacetime that covers the whole spacetime is
\beq
ds^2 = H^{-2}\sec^2\tau (-d\tau^2 + d\chi^2 + \sin^2\chi d\Omega^2)\,,
\eeq
where $|\tau| < \pi/2$ and $0 \leq \chi \leq \pi$, and where $d\Omega^2$ is the metric on the unit $2$-sphere ($S^2$).  The
electromagnetic field from charges $q$ at the North Pole ($\chi=0$) and $-q$ at the South Pole ($\chi=\pi$) can
be
\beqa
A_\tau & = & - \frac{q}{4\pi}\frac{\cos\tau\cot \chi}{\cos\tau+\sin\chi}\,,\label{AHfieldtau}\\
A_\chi & = & - \frac{q}{4\pi}\frac{\sin\tau}{\cos\tau+\sin\chi}\,. \label{AHfieldchi}
\eeqa
In Ref.~\cite{Higuchi:2008fu} this field was reproduced using Eq.~(\ref{AHreproduce}) from the initial data on
past infinity and the current $J^a(x)$
corresponding to the charges at the North and
South Poles in the Feynman gauge ($\zeta=1$).

We first consider the source field due to a charge $q$ at the North Pole.
The causal future  of the North Pole, with the condition $\tau > \chi-\pi/2$, can be
covered by the coordinates $(\lambda,r)$ (and the angular coordinates on $S^2$) given by
\beqa
H\lambda & = & \frac{\cos\tau}{\cos\chi + \sin\tau}\,,\\
Hr & = & \frac{\sin\chi}{\cos\chi+\sin\tau}\,.
\eeqa
The metric in these coordinates is
\beq
ds^2 = \frac{1}{H^2\lambda^2}(-d\lambda^2 + dr^2 + r^2d\Omega^2)\,. \label{AHmetric}
\eeq
The conformal time $\lambda$ decreases toward the future from $\infty$ to $0$.
The North Pole is at $r=0$.  The electromagnetic potential given by Eqs.~(\ref{AHfieldtau})
and (\ref{AHfieldchi}) in this coordinate system is
\beqa
    A_{\lambda} &=& \frac{q}{4\pi}\(\frac{1}{r}-\frac{1}{\lambda+r}\)\,, \label{AHnewcolambda}\\
    A_r &=& -\frac{q}{4\pi(\lambda+r)}\,. \label{AHnewcor}
\eeqa

Following Ref.~\cite{Allen:1985wd}, we define for spacelike separated points $x$ and $x'$
\beq
z \equiv [1+\cos H\mu(x,x')]/2\,,
\eeq
where $\mu(x,x')$ is the geodesic distance between
$x$ and $x'$.  With the notation $x=(\lambda,\mathbf{x})$, $x'=(\lambda',\mathbf{x}')$ we have (see e.g.~Eq.~(4.28) of Ref.~\cite{Higuchi:2008fu})
\beq
\cos H\mu(x,x')= \frac{\lambda^2 + \lambda^{\prime 2} - \|\mathbf{x}-\mathbf{x}'\|^2}{2\lambda\lambda'}\,.
\eeq
This formula allows one to extend the variable $z$ to the cases where $x$ and $x'$ are not spacelike separated.

One can readily find the Feynman propagator for the electromagnetic potential using the method of Allen and
Jacobson~\cite{Allen:1985wd} as
\beq
Q_{ab'}(x,x') = Q^{FG}_{ab'}(x,x') + (\zeta-1)\nabla_a \nabla_{b'}\tilde{\Delta}(x,x')\,,
\eeq
where
\beqa
\frac{\partial\ }{\partial z}
\tilde{\Delta}(x,x')
& = & \frac{H^2}{16\pi^2}\left[\frac{1}{1-z+i\epsilon}-\frac{1}{3z} \right. \nonumber \\
&& \ \ \ \ \ \ \left. - \frac{2z+1}{3z^2}\log(1-z+i\epsilon)\right]\,.
\eeqa
(Since $\nabla_a \tilde{\Delta} = \nabla_a z\partial\tilde{\Delta}/\partial z$, we only need the $z$-derivative
of $\tilde{\Delta}$.)
The retarded Green's function is found as the discontinuity across the cuts from $z=-\infty$ to $z=-1$ and
from $z=1$ to $z=\infty$ on the
complex $z$-plane (see, e.g.~Eq.~(4.10) of Ref.~\cite{Higuchi:2008fu}) as
\beq
G_{ab'}(x,x') = G^{FG}_{ab'}(x,x') + (\zeta -1)\tilde{G}_{ab'}(x,x')\,,  \label{AHretard}
\eeq
where $G^{FG}_{ab'}(x,x')$ is the retarded Green's function in the Feynman gauge~\cite{Higuchi:2008fu,Allen:1985wd} and
\beq
    \tilde G_{ab'}(x,x') = \theta(\lambda'-\lambda)\left[\sigma(z)H^2g_{ab'} + 4\gamma(z)\nabla_a
    z\cdot\nabla_{b'}z\right] \,,
\eeq
with
\beqa
    \sigma(z) &=& \frac{1}{16\pi}\[\delta(1-z) + \frac{2z+1}{3z^2}\theta(z-1)\] \,, \\
    \gamma(z) &=& \frac{1}{16\pi}\[\delta(1-z) - \frac{1}{2}\delta'(1-z) - \frac{1}{6z^3}\theta(z-1)\] \,. \nol \\
\eeqa
The bi-vector $g_{ab'}(x,x')$~\cite{Allen:1985wd} is expressed as~\cite{Higuchi:2008fu}
\beq
g_{ab'} = \frac{1}{H^2}
\left(\d_a \d_{b'}\cos H\mu - \frac{1}{2z}\d_a \cos H\mu\cdot \d_{b'}\cos H\mu\right).
\eeq
If points $x$ and $x'$ are spacelike separated, then the bi-vector $g_{ab'}$ is the parallel propagator along the
geodesic between these points.

Now, we use the retarded Green's function (\ref{AHretard}) to find the source field $A^{(S)}(x)$ given by
Eq.~(\ref{AHsource}) due the charge $q$ at the North Pole $r=0$.  We let the charge be present only for
$\lambda < \lambda_0$ and let $\lambda_0\to \infty$ at the end.
The contribution from $G_{ab'}^{FG}(x,x')$ in Eq.~(\ref{AHretard}) is given by Eqs.~(4.34) and (4.35) of
Ref.~\cite{Higuchi:2008fu}.  Here, we calculate the contribution from the second term in Eq.~(\ref{AHretard}) denoted
by $\tilde{A}^{(S)}_a(x)$.
If we write
\beq
\tilde{G}_{ab'}=\tilde{G}^{(\delta)}_{ab'}\delta(1-z)
+ \tilde{G}^{(\delta')}_{ab'}\delta'(1-z)
+ \tilde{G}^{(\theta)}_{ab'}\theta(z-1)\,,
\eeq
we have
\begin{widetext}
\beq
    \tilde{A}^{(S)}_a(x)
    = -q(\zeta-1)\theta(\lambda_0-\lambda-r)
    \left\{\left. \frac{2\lambda\lambda'}{r}
    \(\tilde{G}^{(\delta)}_{a\lambda'}
    + \frac{\d}{\d\lambda'}\[\(\frac{\d z}{\d\lambda'}\)^{-1}\tilde{G}^{(\delta')}_{a\lambda'}\]
    \)\right|_{\lambda'=\lambda+r}
    + \int_{\lambda+r}^{\infty}d\lambda'\tilde{G}^{(\theta)}_{a\lambda'} \right\}_{r'=0} \,.
\eeq
\end{widetext}
We find that the only non-vanishing component is
\beq
    \tilde{A}^{(S)}_{\lambda} = \frac{q}{12\pi\lambda}\(\zeta-1\)\theta(\lambda_0-\lambda-r) \,. \label{AHsourceterm}
\eeq

Next, we calculate the initial field using the coordinates covering the causal \emph{past} of
the North Pole with the condition $\tau < \pi/2 - \chi$.  Note that its past boundary is past infinity
$\tau = -\pi/2$.  This part of the spacetime is covered by the coordinates $(\hat\lambda,\hat{r})$ defined by
\beqa
H\hat\lambda & = & \frac{\cos\tau}{\cos\chi - \sin\tau} = \frac{\lambda}{H(\lambda^2-r^2)}\,, \label{AHlam}\\
H\hat{r} & = & \frac{\sin\chi}{\cos\chi-\sin\tau}=\frac{r}{H(\lambda^2-r^2)}\,. \label{AHr}
\eeqa
The equalities relating $(\hat\lambda,\hat{r})$ and $(\lambda,r)$ are valid only in the overlapping region
with $\chi - \pi/2 < \tau < \pi/2-\chi$.  The metric in coordinates $(\hat{\lambda},\hat{r})$ is given by
Eq.~(\ref{AHmetric}) with $\lambda$ and $r$ replaced by $\hat{\lambda}$ and
$\hat{r}$, respectively. (The North Pole is again at $\hat{r}=0$.)
The conformal time $\hat{\lambda}$ increases from $0$ to $\infty$,
and the surface $\hat{\lambda}=0$ is past infinity.  In finding the initial field, we let the initial surface
$\Sigma$ be the $\hat{\lambda}=\hat{\lambda}'(={\rm const})$ surface and let $\hat{\lambda}'\to 0$ at the end.
The field given by Eqs.~(\ref{AHfieldtau}) and (\ref{AHfieldchi}) is expressed in coordinates $(\hat\lambda,\hat{r})$
on the initial surface $\hat{\lambda}=\hat{\lambda}'$ (with $\hat{r}=\hat{r}'$) as
\beqa
A_{\hat{\lambda'}} & = & - \frac{q}{4\pi}\left( \frac{1}{\hat{r}'} - \frac{1}{\hat{r}' +\hat{\lambda}'}\right)\,,
\label{AHlowlam}\\
A_{\hat{r}'} & = & \frac{q}{4\pi}\frac{1}{\hat{r}'+\hat{\lambda}'}\,, \label{AHlowr}
\eeqa
with the angular components vanishing (see Eqs.~(4.39) and (4.40) in Ref.~\cite{Higuchi:2008fu}).

Now, it can be shown that $\D^{\left[a'\right.}\tilde{G}_a{}^{\left.b'\right]}=0$, and
$\nabla^{c'}G^{FG}_{ac'} = \nabla^{c'}\tilde{G}_{ac'}$. (That is, the retarded Green's function is divergence-free
in the Landau gauge $\zeta=0$.)  This equality can be used to show that
\beq
    \left. {(L_\pi G)_a}^{a'b'} = {(L_\pi G^{FG})_a}^{a'b'} \right|_{\zeta=1}\,.
\eeq
Thus, the second term in Eq.~(\ref{AHinitial}) is $\zeta$-independent.  This implies that the additional contribution to
the initial field is
\beq
\tilde{A}^{(I)}_a(x) = (\zeta-1)\int_{\Sigma}d\Sigma_{a'}\pi^{a'b'}(x')\tilde{G}_{ab'}(x,x')\,.
\eeq
The field $\pi^{a'b'}(x')$ is defined by Eq.~(\ref{AHpi}).
The additional contribution to the initial field in the causal past of the North Pole thus obtained is
\beqa
    \tilde{A}^{(I)}_{\hat\lambda} &=& \frac{q}{12\pi}\(\zeta-1\)
    \(\frac{1}{\hat\lambda+\hat{r}}+\frac{1}{\hat\lambda-\hat{r}}-\frac{1}{\hat{\lambda}}\) \nol \\
    && \times\theta(\hat\lambda-\hat\lambda'-\hat{r}) \,, \\
    \tilde{A}^{(I)}_{\hat{r}} &=& \frac{q}{12\pi}\(\zeta-1\)
    \(\frac{1}{\hat\lambda+\hat{r}}-\frac{1}{\hat\lambda-\hat{r}}\) \nol \\
    && \times\theta(\hat\lambda-\hat\lambda'-\hat{r}) \,,
\eeqa
and all other components vanish.  Notice that $\tilde{A}_a^{(I)} = 0$ if $\hat{\lambda}-\hat{r} < \hat{\lambda}'$,
i.e. if the point is not in the causal future of the charge at the North Pole.  Note also that the region covered
by the coordinates $(\hat\lambda,\hat{r})$ does not intersect the causal future of the South Pole, and hence that in this
region the source field from the charge at the South Pole vanishes.

Let us present the source and initial fields by adding together
those in the Feynman gauge found in Ref.~\cite{Higuchi:2008fu} and the additional contributions due to the change
in the gauge parameter found in this paper.
The source field is
\beqa
    A^{(S)}_{\lambda} &=& \frac{q}{4\pi}\(\frac{1}{r}-\frac{1}{\lambda+r}+\frac{\zeta}{3\lambda}\)\,,\label{AHfinal1}\\
    A^{(S)}_r &=& -\frac{q}{4\pi(\lambda+r)}\,,
\eeqa
where we have let $\lambda_0\to \infty$.  The initial field is
\beqa
    A^{(I)}_{\hat\lambda} &=& -\frac{q}{4\pi}\left(\frac{1}{\hat r}-\frac{1}{\hat r+\hat\lambda}\right)
    \theta(\hat r-\hat\lambda) \nol \\
    && - \frac{q\zeta}{12\pi}\left(\frac{1}{\hat\lambda}-\frac{1}{\hat\lambda+\hat r}-\frac{1}{\hat\lambda-\hat r}\right)
    \theta(\hat\lambda-\hat r) \,, \ \ \ \ \\
    A^{(I)}_{\hat r} &=& \frac{q}{4\pi(\hat r+\hat\lambda)}\theta(\hat r-\hat\lambda) \nol \\
    && + \frac{q\zeta}{12\pi}\left(\frac{1}{\hat\lambda+\hat r}-\frac{1}{\hat\lambda-\hat r}\right)
    \theta(\hat\lambda-\hat r) \,, \label{AHfinal4}
\eeqa
where we have let $\hat\lambda'\to 0$.  The initial field in the overlapping region
can be written in coordinates $(\lambda,r)$, using Eqs.~(\ref{AHlam}) and (\ref{AHr}), as
\beqa
\left.\tilde{A}_\lambda^{(I)}\right|_{\hat\lambda>\hat{r}} & = & -\frac{q\zeta}{12\pi\lambda}\,,\\
\left.\tilde{A}_r^{(I)}\right|_{\hat\lambda>\hat{r}} & = & 0\,.
\eeqa
Thus, the sum $A_a^{(S)}+A_a^{(I)}$ is $\zeta$-independent and reproduces the field configuration
given by Eqs.~(\ref{AHfieldtau}) and (\ref{AHfieldchi}) in the region with $\tau < \pi/2-\chi$, i.e. in
the causal past of the North Pole.  It is clear that this conclusion holds for the causal past of the
South Pole with $\tau < \chi-\pi/2$.
Hence, we can conclude that Eq.~(\ref{AHreproduce}) reproduces the correct field configuration for
$\tau < 0$.  Then by using the uniqueness of the solution to the field equations for the electromagnetic potential
for given initial data on the Cauchy surface $\tau = -\pi/4$, say,
we conclude that Eq.~(\ref{AHreproduce}) reproduces the correct field over the whole spacetime
for any $\zeta$, extending the result for $\zeta=1$ in  Ref.~\cite{Higuchi:2008fu}.

\acknowledgments

The work of J.R.N. was supported in part by
Nuffield Science Bursary URB/35537 in July 2008.

\end{document}